\RequirePackage{tikz}
\documentclass[pdflatex,sn-aps,iicol,a4paper]{sn-jnl}

\usepackage{caption}
\usepackage{float}
\usepackage{subcaption}
\usepackage{commath}
\usepackage[T1]{fontenc}
\usepackage[utf8]{inputenc}
\usepackage{xspace}
\usepackage{textpos}

\def\eg{\textit{e.g.}\xspace}
\def\mrm{\mathrm}
\def\epem{$e^+e^-$\xspace}
\def\ep{$e\mathrm{p}$\xspace}

\def\pythia{\textsc{Pythia}8\xspace}
\def\angantyr{\textsc{Angantyr}\xspace}
\def\pp{\ensuremath{\mathrm{pp}}\xspace}
\def\pPb{\ensuremath{\mathrm{pPb}}\xspace}

\def\qcdcr{QCDCR\xspace}
\def\eg{\emph{e.g.}\xspace}

\def\cf{\emph{c.f.}\xspace}

\def\setting#1{\texttt{\footnotesize #1}}

\jyear{2023}%

\begin{document}
\title[~]{The dynamic hadronization of charm quarks in heavy-ion collisions}

\author[1]{\fnm{Christian} \sur{Bierlich}}\email{christian.bierlich@hep.lu.se}

\author[1]{\fnm{G{\"o}sta} \sur{Gustafson}}\email{gosta.gustafson@hep.lu.se}

\author[1]{\fnm{Leif} \sur{L{\"o}nnblad}}\email{leif.lonnblad@hep.lu.se}

\author[1,2]{\fnm{Harsh} \sur{Shah}}\email{harsh.shah@hep.lu.se, harsh.shah@ifj.edu.pl}

\affil[1]{\orgdiv{Dept.\ of Physics}, \orgname{Lund University}, \orgaddress{\street{S{\"o}lvegatan\ 14A}, \city{Lund}, \postcode{SE-223~62}, \country{Sweden}}}
\affil[2]{\orgname{Institute of Nuclear Physics, Polish Academy of Sciences, Cracow},\country{Poland}}

\abstract{The \pythia/\angantyr~model for heavy ion collisions was recently updated with a mechanism for \textit{global colour reconnection}. The colour reconnection model used is QCD colour algebra inspired and enhances baryon production due to the formation of string junctions. In this paper, we present updates to the junction formation and string fragmentation mechanisms, connected to heavy quark fragmentation. This allows for the simulation of heavy quark fragmentation, using junction formation, in heavy ion collisions. The framework is validated for proton collisions, and we show results for charm baryon production in proton-lead collisions.}


\keywords{Heavy flavour, Hadronization, Heavy-ion collisions}



\maketitle

\section{Introduction}
\label{sec:intro}

In high-energy particle collisions, hadrons with heavy quark content, are a uniquely versatile probe of fragmentation dynamics. Their defining feature, a charm ($\mrm{c}$) or bottom ($\mrm{b}$) flavoured quark, cannot originate from the hadronization process but must be created either in the hard process or in the parton shower, both calculable with perturbative techniques.

As opposed to the even heavier top ($\mrm{t}$) quark, hadrons containing $\mrm{c}$- and $\mrm{b}$--type quark content, are still understood to fragment through the same mechanisms as their light counterparts, the $\mrm{u}$, $\mrm{d}$, and $\mrm{s}$ quarks. When comparing experimental data to theory, two quite different (and thus complementary) techniques are used: The factorisation approach and the route taken by Monte Carlo event generators.
In the factorisation approach \cite{Collins:1985gm, Collins:1989gx}, the cross-section is separated into a convolution of three factors: 1) a Parton Distribution Function (PDF) of the incoming hadron, 2) the parton level hard scattering cross-section, where state-of-the-art calculations today are next-to-leading-order (NLO) in the strong coupling ($\alpha_s$) (see \eg \cite{Nason:1987xz, Nason:1989zy, Mangano:1991jk}) often with next-to-leading-log (NLL) resummation techniques applied as well,
 such as \eg GM--VFNS \cite{Helenius:2018uul} or FONLL \cite{Cacciari:1998it, Cacciari:2012ny}, and finally 3) fragmentation functions, analytical expressions fitted to \epem and \ep data \cite{Belle:2005mtx, Kneesch:2007ey} giving differential probabilities for the charm quark to fragment to various hadron species. It has been known at least since SPS \cite{WA82:1993ghz} that the underlying assumption of independent fragmentation does not hold, but it has generally been assumed that universal fragmentation functions can be applied across systems, when studying inclusive quantities, such as total charm hadron yields per event. Recent work by the ALICE collaboration \cite{ALICE:2020wfu, ALICE:2021dhb,ALICE:2021rzj, ALICE:2023jgm} has, however, clearly shown that fragmentation functions tuned to \epem and \ep, cannot describe the fragmentation of charm into baryons in \pp. 

In the Monte Carlo event generator approach, as used in \eg~\pythia~\cite{Bierlich:2022pfr}, PDFs are still used to extract the
participating partons from the colliding nucleons. But where the focus in the factorisation approach tends to be more directed towards formal precision in the calculation of the hard scattering, the focus in the Monte Carlo generators is more towards coherent modelling of both perturbative and non-perturbative aspects, such as hadronization. Once the total amount of charm quarks present in the event is determined by means of a leading order calculation, plus parton shower \cite{Norrbin:1998bw,Norrbin:2000zc}, the amount of hadron species, is determined by the dynamical fragmentation model, the Lund string model \cite{Andersson:1983ia}, and its extensions. This makes charm hadrons very well suited for studies of dynamical hadronization models. For charm baryon production in particular, the so-called QCD colour reconnection (CR) model \cite{QCDCR} in \pythia has gained a lot of attention, due to its ability to correctly reproduce the $\Lambda^+_c$ yield and $\Lambda^+_{c}/D^0$ ratio as a function of $p_\perp$ in $\pp$ collisions at various collision energies at LHC \cite{ALICE:2020wfu, ALICE:2021rzj, ALICE:2023jgm}. However, the predicted production rates of $\Xi_c$ and $\Omega_c$ baryons are still undershooting data, even with the \qcdcr model \cite{ALICE:2021bli, ALICE:2021psx, ALICE:2022cop}. Furthermore, the model has, until recently not been usable for heavy ion collisions.

One of the key aspects of the \qcdcr model is the formation of junction-like configurations between two or three colour dipoles.
These junction systems contribute to baryon production in addition to the baryons produced during the string fragmentation in \pythia.
We have recently improved the junction fragmentation for the low-energy junction systems and extended the \qcdcr model with a spatial constraint \cite{Lonnblad:2023stc}.
As a result, the \qcdcr model can be used as a global CR model for heavy-ion collision simulations in the \angantyr model \cite{Angantyr, Lonnblad:2023stc}.

In this paper, we further improve the junction formation and fragmentation for the colour dipoles containing heavy quarks. We use \pp collisions to validate the framework and show for the first time how \pPb collisions generated with \angantyr + \qcdcr, give a satisfactory description of $\Lambda^+_c$ production.
We show results primarily for charm baryons, but a similar outcome can be expected for the bottom quark containing baryons as well.
We present the results using the upgraded \qcdcr model from \cite{Lonnblad:2023stc} with the new changes we have made in this paper.

We first provide an overview of the \angantyr model for heavy-ion event simulation in \pythia in the next section.
In section \ref{sec:cHad} we discuss the perturbative production of the charm quarks, and in section \ref{subsec:HadronProd}, we show the non-perturbative aspects of the charm hadrons production.
We also discuss the changes we have made in junction formation and fragmentation.
Finally, results for charm hadron production in \pp and \pPb are shown in section \ref{sec:results}.

\section{Heavy ion collisions with the \angantyr model}

The \angantyr model \cite{Angantyr,Bierlich:2016smv} is an extension of \pythia to simulate heavy-ion collision events without assuming the creation of a Quark--Gluon plasma.
It uses a modified Glauber model \cite{Glauber:1955qq,Miller:2007ri} to obtain the number and types (\eg elastic or inelastic (diffractive or non-diffractive) interactions) of sub-collisions in a heavy-ion collision event.
Based on the number and type of sub-collisions, multiple $\pp$-like collisions are generated and stacked together to produce the heavy-ion event.

The arrangement of the nucleons inside a nucleus is obtained using the Woods-Saxon distribution in the GLISSANDO parametrization \cite{Bozek:2019wyr}.
When nuclei collide with each other at relativistic energies, they are Lorentz contracted. The wave functions of the nucleons inside the nuclei can be treated as frozen at the time of the collision.
This is realized in the so-called Glauber-Gribov \cite{Heiselberg:1991is,Blaettel:1993ah,Alvioli:2013vk} formalism for nucleon wave-function fluctuations and extended it to include cross-section fluctuations in projectile and target nucleons for pA and AA collisions.

Once the types of nucleon-nucleon (NN) sub-collisions are decided, the \angantyr model uses the \pythia model for multiparton interactions to generate respectively non-diffractive, diffractive, and elastic \pp events.
Often it occurs that a nucleon is participating in more than one NN non-diffractive sub-collision. A key feature of the model is the special treatment for nucleons participating in multiple non-diffractive interactions.
Given a single projectile nucleon interacting with several target nucleons, the NN pair with the smallest impact parameter is denoted the ``primary'' non-diffractive sub-collision. The others are denoted ``secondary''. The primary sub-collision is generated as a normal non-diffractive \pp collision, whereas the secondaries are generated as a modified single diffractive collision (see section 5 in reference \cite{Angantyr} for further explanation).
A secondary non-diffractive interaction will be discarded once sufficient energy is no longer available.

There is no interaction between the partons produced in different sub-collisions in the default \angantyr.
All multiple sub-collisions are stacked together at the parton level as colour singlet Lund strings.
Later, the Lund strings are hadronised and produce a heavy-ion collision event.

Recently we have added a global colour reconnection (CR) in \angantyr \cite{Lonnblad:2023stc}.
We have extended the \qcdcr model \cite{QCDCR}, by adding a spatial constraint on the colour dipoles to be colour reconnected.
We stack the colour dipoles from different sub-collisions and use the spatially constrained \qcdcr model such that colour dipoles from nearby sub-collisions can undergo CR.
In this work, we continue to use this upgraded \angantyr set-up to simulate \pPb collision events.

\section{Charm hadron production in \pythia}
\label{sec:cHad}

Since the masses of charm ($\approx$1.5~GeV) and bottom ($\approx$4.8~GeV) are large compared to the light quarks, they will never be produced through the tunnelling mechanism by which the string breaks, but only in the hard process and the parton shower. In sub-section \ref{subsec:hardPshowers} we will briefly review the \pythia formalism for heavy quark production, and in sub-section \ref{subsec:HadronProd} we give an overview of the impact of CR. In the following sub-sections, the modifications relevant to charm production in \pPb will be introduced.

\subsection{Charm quark in hard process and parton shower}
\label{subsec:hardPshowers}

Several different QCD processes in $\pp$ collisions in \pythia can produce heavy quarks.
The leading order (LO) processes like $qq \rightarrow Q\Bar{Q}$ and $gg \rightarrow Q\Bar{Q}$ hard scatterings are the primary processes for heavy quarks production in \pythia. Another source of heavy quark production is weak decays ($Z$ and $W^{\pm}$ bosons decays), Higgs decay, and top and bottom quark decay, though of those, only the latter contributes in any significant amount when considering total charm production down to low $p_\perp$.
Furthermore, parton showers, where initial or final state partons (mostly gluons) produce the heavy quarks by pair creation, flavour excitation or gluon splitting. This is a significant source of charm production, in addition to that produced in the hard scattering. Furthermore, the ``hidden charm'' from the PDF  of one of the colliding beams, may come on a mass shell due to the scattering.
The interaction is like $Qq \rightarrow Qq$ or $Qg \rightarrow Qg$, but since the Q is not a valence quark it has to be produced in pairs by a gluon splitting.

The LO processes have the matrix elements containing the heavy quark mass.
Since quark masses are included, full phase space down to $p_{\perp} \rightarrow 0$ can be populated. For low $p_\perp$ production, however, using the \pythia multiparton interaction framework \cite{Sjostrand:1987su}, which
introduces a general parameter $p_{\perp 0}$, is more suitable, in particular when extending to heavy ion collisions.
The heavy quark masses are an important parameter in the perturbative description of their production.
In \pythia, the default values for the charm and bottom quark masses are set to 1.5~GeV and 4.8~GeV respectively. The masses affect the matrix elements, splitting kernels, and the phase space of the heavy quarks production cross-section.
These values are fitted to $D$-meson production rates. To better fit production rates at LHC in \cite{Bierlich:2023ewv} authors show that a reduced charm quark mass is expected. Following the arguments in \cite{Bierlich:2023ewv} one can also expect a similar correction in the bottom quark mass. We have reduced the charm and bottom quark masses to 1.3~GeV and 4.2~GeV respectively in this paper.

\subsection{Colour reconnection and hadronization}
\label{subsec:HadronProd}

After the multiple parton scatterings and parton showers, outgoing quarks and gluons are connected by strings. We speak of a string connecting a colour and an anti-colour -- either quark and anti-quark or through one or more gluon ``kinks'' -- as a chain of colour dipoles. These colour connections 
are reassigned through colour reconnection (CR) \cite{Sjostrand:1987su, QCDCR} models. The conventional argument, and indeed the logic behind the
default CR model in \pythia, is that while the parton shower generates a colour configuration in the $N_{c}\rightarrow \infty$ limit, nature has 
$N_c = 3$. The choice of specific colour connections for a single event is ambiguous and should therefore be corrected. The calculation itself, however, cannot provide
any guidance as to how to do the reconnection, and one must resort to models. A common feature is a reduction of the so-called $\lambda$-measure, which is an indirect representation of the rapidity span of the colour dipoles, which is again a logarithmic sum of the potential energy of the dipoles, and hence a measure of the number of hadrons produced by the dipoles. Further details in sub-section \ref{subsec:junction-formation}. 
The CR in \pythia helps to reproduce the charged particle multiplicity and the increase in $\langle p_\perp \rangle$ as a function of $(N_{ch})$ distribution as observed in the experiments.

The \qcdcr model \cite{QCDCR} is developed with the idea of applying $\textbf{SU(3)}$ colour algebra on non-correlated colour dipoles before calculating the $\lambda$-measure for the new configurations of the dipoles.
Colour algebra allows the formation of a colour singlet by three colour string pieces being connected to a ``junction'' point (see figure \ref{fig:dipCR}).
The ``string system contains a junction'' (junction system) formed by two or three dipoles is not possible in the earlier case of $N_{c}\rightarrow \infty$ limit.
Hence in the \qcdcr model, the two and three dipoles can have three string pieces that are colour-connected to a single ``junction'' point after the CR.
A junction system produces at least one baryon per junction during the hadronization.
In the \qcdcr model, junctions are always produced as junction and anti-junction pairs and conserve the baryon number.
These baryons (and anti-baryons) are additional baryons due to \qcdcr. 

After the CR, the colour singlet Lund strings hadronised by sequential fragmentation.
The different flavours of quarks and anti-quarks are produced according to the Lund string model \cite{LSM}.
Parameter values are fixed from the model tuning with LEP data \cite{Skands:2014pea,QCDCR,Bierlich:2015rha}.

The sequence of the string breaks decides if a string piece will form a meson or a baryon as a primary hadron.
The Lund string fragmenting into $q \bar{q}$ pairs will produce mesons.
For baryons production, the string has to break into a diquark--anti-diquark pair, where the consecutive string breaks of a $q \bar{q}$ pair on either side of the diquarks will produce a baryon and an anti-baryon.
\pythia uses the ``popcorn mechanism'' \cite{Bar:popcorn}, which includes a probability for a meson production between the baryon and the anti-baryon, and the results of the ``popcorn mechanism'' are supported by the experiments \cite{TPCTwoGamma:1985zxy}.

A $q\Bar{q}$ pair production rate,
\begin{equation}
	\mathrm{d}P \simeq \mathrm{d}^2p_{\perp} \exp \left( -\pi m^2_{\perp}/\kappa \right), \hspace{4mm} m^2_\perp = p^2_\perp + m^2_{q},
  \label{eq:breakup}
\end{equation}
where $\kappa$ is string tension, and $m_\perp$ is transverse mass of the quark with mass $m_{q}$ and opposite transverse momenta $p_\perp$.
The pair production rate is mass-dependent, and it gives an extremely low probability for the production of heavy quarks pair (\eg "charm" pairs) during the string fragmentation.
Therefore all of the heavy quarks are produced either in hard scatterings or in parton showers in \pythia as mentioned in section \ref{subsec:hardPshowers}.
In this paper, we show results for charm baryons only, so we refer to charm quarks as the heavy quarks for the rest of the paper.

The heavy quarks form mesons or baryons depending on which quarks/diquarks are produced next to them during string fragmentation.
The only way the hadronization of the heavy quarks can be influenced is either by modifying fragmentation parameters or by colour reconnection.
LEP data constrains the fragmentation parameters, while in the CR we have some freedom in rearranging the colour connections among the partons. 
Moreover, the \qcdcr model allows junction configurations, which contribute to baryon production.
In such a case the type of partons attached to the junction legs and the choice of junction fragmentation sequence can influence the baryon production.
Moreover, the probability for junction formation increases in a more dense environment like high multiplicity $\pp$ or heavy nuclei collisions.

Until now, no special attention was given if a heavy quark is involved during the junction formation or fragmentation.
In this paper, we improve the junction treatments if one or more heavy quarks are involved.
We discuss these improvements below in detail.

\subsection{The role of junctions}
\label{subsec:role-junctions}
We have made several improvements in the \qcdcr model in
\cite{Lonnblad:2023stc}.  A crucial addition is the impact
parameter-dependent constraint on the colour dipoles to be colour
reconnected, which allowed us to have a global CR among the partons
produced in different sub-collisions in heavy-ion event simulation in
\angantyr. In addition, we implemented a few technical improvements in
the hadronisation of junctions.

In \pythia, there may occur situations when a string system in an
event cannot be hadronised properly. There are several reasons for
such failures, and often they involve junctions. If such a failure
arises, \pythia will throw away the whole event and generate a new
one. In \pp\ collisions such failures are typically rare, but in
heavy-ion collisions there can be very many strings and the failure
rate per event increases. And since the rate is higher for high
multiplicity events (many strings), there is a risk that the overall
multiplicity distributions may be skewed. With the \qcdcr model, the
number of junctions increases, which also increases the failure rate,
and we found in \cite{Lonnblad:2023stc} that the effect on
multiplicity distributions was substantial in the heavy-ion collision, and
even visible in \pp\ collisions.

A majority of the discarded events are found to have at least one
junction system with a very low invariant mass ($\lesssim$1~GeV). We
added a ``junction collapse'' mechanism to hadronize the low-mass
junction systems, which were not treated in \pythia prior to
\cite{Lonnblad:2023stc}.  This ``junction collapse'' mechanism
produces two hadrons from the junction system.  These hadrons can be
two baryons or a meson and a baryon depending on the types of partons
attached to the end of every leg in a junction system.  We have also
introduced an additional trial if the string fragmentation fails to
hadronise a junction system.  In such a scenario, the junction system
is fragmented by the special version of the junction collapse
procedure.

As we mentioned at the beginning of the paper, the conflicting results
from \epem and $\pp$ collisions raise questions about the universality
of charm hadron production. In \pythia, this applies in particular to
charm baryons. We have seen that adding \qcdcr improves the description
for $\Lambda_c$, but for heavier charmed baryons there is still a
problem.
Since the additional junctions from the \qcdcr model are responsible
for the increased charm baryon rates, we want to look in more detail
into how junctions involving charm quarks are handled there, and also
how they are treated in the subsequent string fragmentation.

\subsubsection{Junction formation}
\label{subsec:junction-formation}
\begin{figure*}
\centering
\begin{tikzpicture}

\draw (-1,0) -- (2,0);
\filldraw [gray] (-1,0) circle (2pt) node[left]{$q$};
\filldraw [gray] (2,0) circle (2pt) node[right]{$\bar{q}$};
\draw (-1,1) -- (2,1);
\filldraw [gray] (-1,1) circle (2pt) node[left]{$q$};
\filldraw [gray] (2,1) circle (2pt) node[right]{$\bar{q}$};

\draw [-stealth](3.2,0.5) -- (4.8,0.5);

\draw (6,0) -- (6.5,0.5);
\filldraw [gray] (6,0) circle (2pt) node[left]{$q$};
\draw (6,1) -- (6.5,0.5);
\filldraw [gray] (6,1) circle (2pt) node[left]{$q$};

\draw (6.5,0.5) -- (8.5, 0.5);

\draw (9,0) -- (8.5,0.5);
\filldraw [gray] (9,0) circle (2pt) node[right]{$\bar{q}$};
\draw (9,1) -- (8.5,0.5);
\filldraw [gray] (9,1) circle (2pt) node[right]{$\bar{q}$};
\node[] at (4,-0.5) {(a)};

\draw (-2,-1) -- (10, -1);
\draw (-1,-2) -- (2,-2);
\filldraw [gray] (-1,-2) circle (2pt) node[left]{$q$};
\filldraw [gray] (2,-2) circle (2pt) node[right]{$\bar{q}$};
\draw (-1,-3) -- (2,-3);
\filldraw [gray] (-1,-3) circle (2pt) node[left]{$q$};
\filldraw [gray] (2,-3) circle (2pt) node[right]{$\bar{q}$};
\draw (-1,-4) -- (2,-4);
\filldraw [gray] (-1,-4) circle (2pt) node[left]{$q$};
\filldraw [gray] (2,-4) circle (2pt) node[right]{$\bar{q}$};

\draw [-stealth](3.2,-3) -- (4.8,-3);

\draw (6,-2) -- (7,-3);
\filldraw [gray] (6,-2) circle (2pt) node[left]{$q$};
\draw (6,-3) -- (7,-3);
\filldraw [gray] (6,-3) circle (2pt) node[left]{$q$};
\draw (6,-4) -- (7,-3);
\filldraw [gray] (6,-4) circle (2pt) node[left]{$q$};

\draw (9,-2) -- (8,-3);
\filldraw [gray] (9,-2) circle (2pt) node[right]{$\bar{q}$};
\draw (9,-3) -- (8,-3);
\filldraw [gray] (9,-3) circle (2pt) node[right]{$\bar{q}$};
\draw (9,-4) -- (8,-3);
\filldraw [gray] (9,-4) circle (2pt) node[right]{$\bar{q}$};
\node[] at (4,-4) {(b)};

\end{tikzpicture}
\caption{Illustration of colour reconnections forming junctions. Two dipoles can form a colour connected junction--anti-junction system (a), and three can form two separate (anti-) junction systems (b).}
\label{fig:dipCR}
\end{figure*}
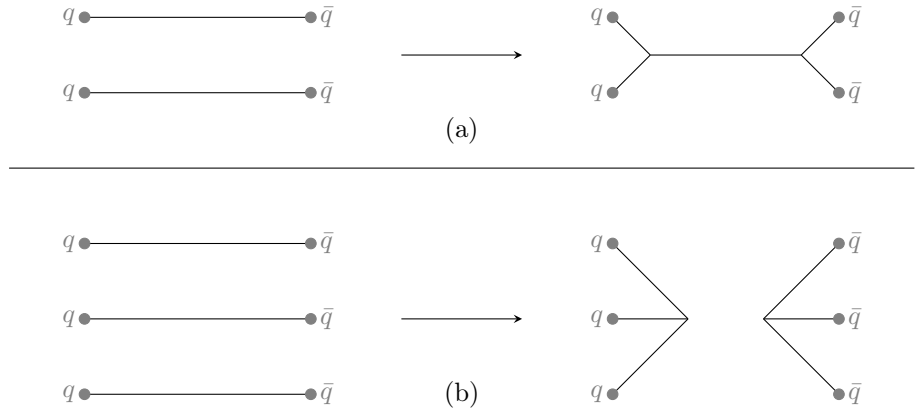

In \pp collisions, junctions are normally only formed in the treatment
of the proton remnants, when more than one valence quark undergo
scattering in the multiple parton interactions machinery, but such junctions mainly influence baryon production in the forward
rapidity region.  The junction formation due to colour reconnection is
unique to the \qcdcr model.  The QCD colour algebra-based reconnection
treatment includes the colour connections beyond the leading colour
approximations. This means that besides the case where two
uncorrelated dipoles having the exact same colour state can ``swing''
so that the coloured parton in one dipole becomes colour connected
with the anti-colour of the other, and vice versa, there can also be
reconnections between dipoles that have different colour states.  In
this way, the partons in two or three colour dipoles can become
colour-connected to junction points (as shown in figure
\ref{fig:dipCR})
with a certain probability that they are carrying the right colour
charges. Each of the junction legs has to have a different colour
charge so that the junction system becomes a colour singlet.

After all possible reconnections are tabulated, the \qcdcr model will
order them so that the reconnection reduces the string
lengths the most, as defined by the $\lambda$-measure are performed
first. For dipoles between two partons the $\lambda$-measure in
the model is given by\footnote{The infrared $\lambda$-measure for a string in Ref. \cite{Andersson:1988ee}, is in the \qcdcr model approximated by the sum of these contributions from the individual dipoles.}
\begin{equation}
  \lambda = \ln \left( 1 + \frac{\sqrt{2}E_1}{m_0} \right) + \ln \left( 1 + \frac{\sqrt{2}E_2}{m_0} \right),
  \label{eq:2}
\end{equation}
where the energies $E_i$ are given in the dipole's rest frame, and
$m_0$ is a tunable parameter. For a dipole connecting a parton to a
junction, the model similarly defines the $\lambda$-measure as
\begin{equation}
    \lambda_j =  \ln \left( 1 + \frac{\sqrt{2}E}{m_j} \right),
    \label{eq:3}
\end{equation}
where $E$ is the energy of the parton given in the junction rest
frame\footnote{In the junction rest frame, the angles between all momenta of the connected partons are $120^\circ$.} and $m_j$ is a tunable parameter not necessarily the same as $m_0$.

As discussed in the introduction the $\lambda$-measure is an estimate of the rapidity range for the hadrons in the string breakup.
The definitions in equations (\ref{eq:2} and \ref{eq:3}) are well motivated for light quarks and massless gluons.
However, for a string piece connected to a heavy quark, these expressions are not good estimates of the rapidity range.
In this case, we instead use the rapidity of the heavy quark in the rest frame of the junction:

\begin{equation}
    \lambda_{HQ} = \frac{1}{2} \log \left( \frac{E + p}{E - p} \right).
\end{equation}
Here $E$ and $p$ are the energy and momentum of the heavy quark in the junction rest frame.
The $\lambda_{HQ}$ will give a lower value than $\lambda_j$ for heavy quarks, especially for small $p$. Hence with this new change, we enhance the possibility for a heavy quark to be part of a junction system during CR in \pythia. We note that there is no need for the
parameter $m_j$ to set the scale in $\lambda_{HQ}$ since the quark mass does that for us. Also, the ``1+'' in the logarithm, which protects the $\lambda$ from becoming negative is also not needed.

\subsubsection{Junction fragmentation}
\label{sec:juncfrag}

After the Colour Reconnections, the colour strings will undergo string fragmentation.
In the \qcdcr model in \pythia, the three junction legs are treated separately according to the following steps.
\begin{itemize}
\item A few attempts are made to move the junction system to the junction rest frame.
\item If the algorithm fails to obtain the junction rest frame, then the junction system fragments in the centre of the mass frame of the junction system.
\item Once the frame is found, the summed energy of the partons on each junction leg is calculated in that frame and the junction legs are tagged as low-, middle-, and high-energy legs.
\item The low-energy leg is fragmented first. A fictitious particle is
  assumed on the opposite side of the junction point for the given
  junction leg, and the string fragments from the endpoint towards the
  junction point until a parton closest to the junction point is left
  on the junction leg.
\item Similarly the middle-energy leg is fragmented.
\item A diquark is formed by combining the flavour and momenta of the
  two partons left on the low and middle legs closest to the junction
  point.
\item The diquark is connected at one end of the high-energy leg, the
  junction no longer exists and the string is fragmented via the usual
  string fragmentation mechanism.
\end{itemize}

Finding the junction frame for three $\textit{massless partons}$ is trivial, but
as soon as one or more of them are massive, the process does not always converge to a stable solution, because such a solution does not exist.
(Also in the case where a junction leg has a long chain of gluons, the proper frame can be difficult to find.)

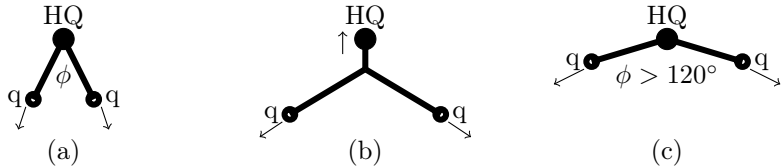
\begin{figure*}[h!]
\centering
\begin{tikzpicture}
\filldraw [black] (-3,-1) circle (4pt) node[above]{HQ};
\path (-3,-1.5) node {$\phi$};
\draw [line width = 0.8mm] (-3,-1) -- (-3.4, -1.8) circle (2pt) node [left] {q};
\draw [line width = 0.8mm] (-3,-1) -- (-2.6, -1.8) circle (2pt) node [right] {q};
\draw [->] (-3.5, -1.9) -- (-3.6, -2.2);
\draw [->] (-2.5, -1.9) -- (-2.4, -2.2);
\path (-3, -2.5) node {(a)};

\filldraw [black] (1,-1) circle (4pt) node[above]{HQ};
\draw [->] (0.7, -1.2) -- (0.7, -0.9);
\draw [line width = 0.8mm] (1,-1) -- (1,-1.4);
\draw [line width = 0.8mm] (1,-1.4) -- (0, -2) circle (2pt) node [left] {q};
\draw [line width = 0.8mm] (1,-1.4) -- (2, -2) circle (2pt) node [right] {q};
\draw [->] (-0.1, -2.1) -- (-0.4, -2.3);
\draw [->] (2.1, -2.1) -- (2.4, -2.3);

\path (1, -2.5) node {(b)};

\filldraw [black] (5,-1) circle (4pt) node[above]{HQ};
\path (5,-1.5) node {$\phi > 120^{\circ}$};
\draw [line width = 0.8mm] (5,-1) -- (4, -1.3) circle (2pt) node [left] {q};
\draw [line width = 0.8mm] (5,-1) -- (6, -1.3) circle (2pt) node [right] {q};
\draw [->] (3.9, -1.4) -- (3.5, -1.6);
\draw [->] (6.1, -1.4) -- (6.5, -1.6);

\path (5, -2.5) node {(c)};

\end{tikzpicture}
\caption{(a) A representation of a junction system with a heavy quark (denoted with HQ) and two light partons denoted with (q), (the arrows show their momentum vectors) with an angle $\phi$ between them, (b) a junction rest frame system for the same configuration, and (c) a scenario where the heavy quark coincides with the junction point and the angle between two light partons is greater than $120^{\circ}$.}
\label{fig:HQrest}
\end{figure*}

Figure \ref{fig:HQrest} (a) illustrates a system with a heavy quark and two light colour charges (quarks or gluons) in the initial rest frame of the heavy quark.
If the angle $\phi$ between the light charges is smaller than $120^{\circ}$, there is always a frame, in which a junction is at rest as in figure \ref{fig:HQrest} (b).
The massive quark moves more slowly, and the corresponding string piece is shorter.
For $\phi = 120^{\circ}$ this length goes to zero, and for $\phi > 120^{\circ}$ the junction coincides with the heavy quark, see figure \ref{fig:HQrest} (c).
In this case, we find that it is most natural to hadronize the system in the rest frame of the heavy quark.
We have implemented this in \pythia, and one consequence of this new procedure is that the heavy quark is more likely to be the lowest energy leg, and will in addition not be able to fragment into a heavy meson before being joined into a diquark and then ending up in a baryon.
(We note that in this situation the \pythia fragmentation system instead hadronizes three strings connected at the centre in the rest frame of the whole junction system.
This reduces the probability of producing a heavy baryon and overestimates the number of produced hadrons.)

We have made one more change in \pythia to enhance the chance of a heavy quark ending up in a baryon, and that is to change the ordering of the junction legs. Instead of taking the leg with the lowest summed energy of the connected partons, we use the sum of absolute spatial momentum instead. In analogy with the change in the $\lambda$-measure in the \qcdcr, this will more closely correspond to how long the
actual string is, and will more often put the leg with a heavy quark among the two legs that are fragmented first. Again this improves the chances that the heavy quark ends up in a baryon.

Besides changing the actual algorithms in \pythia and in the \qcdcr
model, we have also investigated some of the parameters that can
affect the production of charmed baryons.
In the diquark formation by combining the two quarks from the low- and
middle-energy legs, the spin assignment is done by a set of
parameters\footnote{The \pythia parameter used to set these values is
  \setting{StringFlav:probQQ1toQQ0join}.}  suppressing the expected
ratio of 3 spin-1 vs.\ spin-0 states. The default values in \pythia
are 0.5, 0.7, 0.9, and 1, for the cases where the heaviest quark is
u/d, s, c, and b, respectively, but these are not well constrained by
experimental data. In the so-called mode-0 tune for the \qcdcr model,
these were instead all set at 0.0275, which is close to the more well
constrained value used for the diquark--anti-diquark breakups in a
normal string. There is, however, no reason to expect that these
parameters should be the same, since the formation of diquarks in the
joining of junction legs is very different from the breakup in
strings. And since we know that the \qcdcr model has difficulties in
describing the production of heavier baryon states, we have checked
the effect of raising the values to the default ones in \pythia
also when using \qcdcr.

Since we will here mostly be concerned with charmed baryons that also
includes strange quarks, there are also other effects that can
influence the production. It is well-known that strangeness
enhancement is present not only in heavy-ion collisions but also in
high multiplicity \pp collisions (see, \eg, \cite{ALICE:2016fzo}). In
Lund we have studied the so-called rope hadronisation model
\cite{Bierlich:2014xba,Bierlich:2022oja,Bierlich:2022ned}, where
overlapping strings gives an increase in the string tension,
$\kappa$. This results in an increased probability of strange quarks
in the string breakups (\cf, eq.~\ref{eq:breakup}) and the results are
promising. Our current implementation does not, however, handle
junctions very well, which is why we here have decided to emulate the
effect by increasing the overall relative probability of having
strange quarks in string breakups\footnote{The \pythia parameter for this is called \setting{StringFlav:probStoUD}.} from the default
value of 0.217 to 0.4. The number may seem to be high but since most
of the charmed baryons are produced at high multiplicities, where
there are many dipoles that can reconnect, and hence also many strings
can overlap, we do not consider it to be unreasonably high.

The overall charm content in an event is mainly governed by
perturbative effects, and can be gauged by the rate of the most common
charmed D-mesons, which are reasonably well described by the default
\pythia. With the modifications we have described here, however, a
larger fraction of charm quarks will end up in baryons, reducing the
rate of D-mesons, and we have therefore decided to compensate for this by
increasing the overall charm production by reducing the charm (and
bottom) quark mass i \pythia from the default value of 1.5 (4.8)~GeV
to 1.3 (4.2)~GeV.

\begin{figure*}
\begin{subfigure}{.25\textwidth}
  \includegraphics[width=\linewidth]{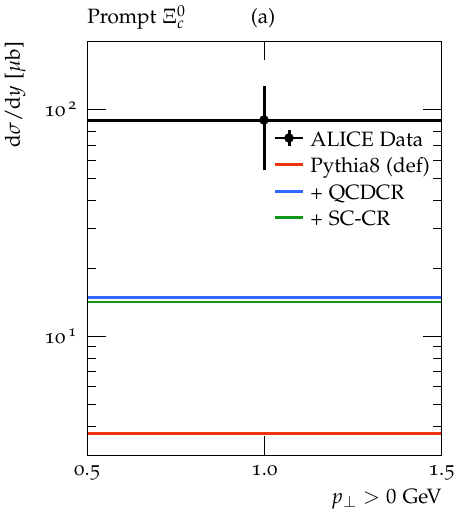}
\end{subfigure}%
\begin{subfigure}{.25\textwidth}
  \includegraphics[width=\linewidth]{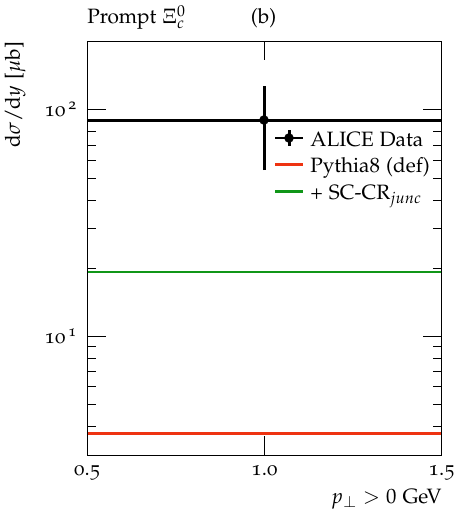}
\end{subfigure}%
\begin{subfigure}{.25\textwidth}
  \includegraphics[width=\linewidth]{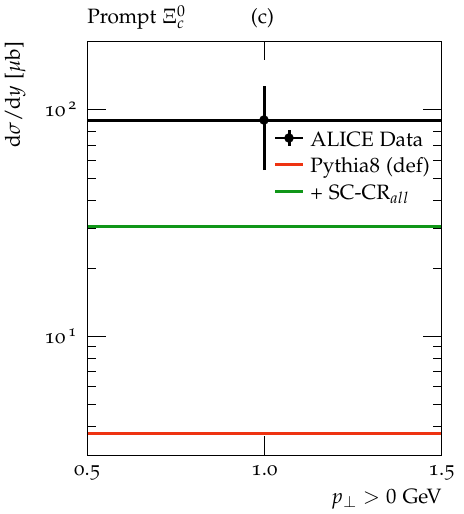}
\end{subfigure}%
\begin{subfigure}{.25\textwidth}
  \includegraphics[width=\linewidth]{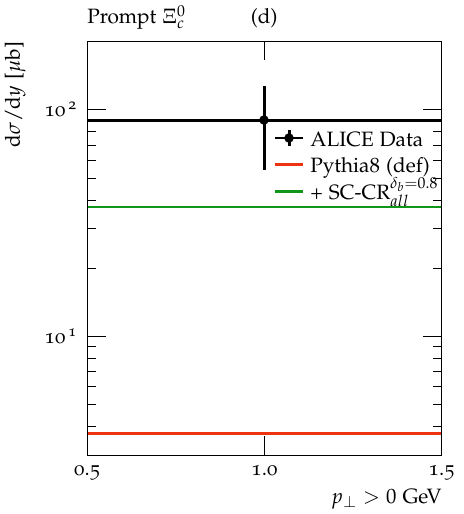}
\end{subfigure}
\caption{Integrated prompt $\Xi^0_c$ cross-section for \pp at
  $\sqrt{s} = 5.02$~TeV, for $\mid y\mid < 0.5$ using different options for
  \pythia compared to ALICE data \cite{ALICE:2021psx}. In all cases,
  we show \pythia default and, from the left, we show (a)~\qcdcr
  (mode-0), and spatially constrained \qcdcr (as SC-CR); (b)~SC-CR
  with corrections in the heavy quarks junction formation and
  fragmentation; (c)~SC-CR will all the new changes from this work;
  and (d)~SC-CR (with all changes) with the $\delta_b$ parameter set
  to 0.8~fm.}
\label{fig:Xic_tests}
\end{figure*}

To get an indication of the overall effects of the changes we have
proposed here, we show in figure \ref{fig:Xic_tests} the rate of direct
production of $\Xi^0_c$ baryons in $\pp$ collisions at
$\sqrt{s} = 5.02$~TeV.  The model results are compared with the ALICE
data \cite{ALICE:2021psx} using the Rivet \cite{Bierlich:2019rhm}
routine called \setting{ALICE\_2021\_I1863039}.  In the left-most
histogram, we show the results of the default \pythia (red line),
\qcdcr (mode-0) (blue line), and spatially constrained \qcdcr (green
line). Here we see clearly the effect of introducing the junction
reconnections in \qcdcr. Our spatially constrained version of the
\qcdcr gives a slightly reduced rate, mainly because of the
constraint, but also because of differently tuned parameters (see
\cite{Lonnblad:2023stc} for details). In the second to the left
histogram, we show the effect of the changes in junction formation and
fragmentation for the SC-CR case, and find an increase of around
35\%. In the third histogram, we have also added the parameter changes
described above and found an additional increase of almost 60\%, giving
an almost doubled rate compared to the default SC-CR, and a
factor 8 more than the default \pythia. We are, however, still far
away from the lower bound of the experimental error bar, and a factor
almost three below the central value.

Finally, in the right-most histogram of figure \ref{fig:Xic_tests}, we
show that if we allow reconnection of dipoles farther separated in the
transverse plane by increasing the spatial constraint ($\delta_b$)
value in the SC-CR model from 0.5~fm to 0.8~fm on top of the other
changes we have made, then we can further enhance the $\Xi^0_c$
baryon's production in \pp collision events.  However, since one
of the aims of this paper is to compare to \pPb data using the
\angantyr model we will in the following keep the tuned value of 0.5~fm,
which we have shown in \cite{Lonnblad:2023stc} gives a more reasonable
description of multiplicities in \pPb.

\section{Results}
\label{sec:results}

In this section, we want to look more in detail at the effects of the
changes we made. We will concentrate on the charmed baryons, but will
also look at non-charmed hyperons. We first look at \pp collisions to
check that we get reasonable results there before we extrapolate the
models to \pPb collisions using \angantyr.

\subsection{Hyperon production in $\pp$ collisions}

\begin{figure*}
\centering
\begin{subfigure}{.43\textwidth}
  \includegraphics[width=\linewidth]{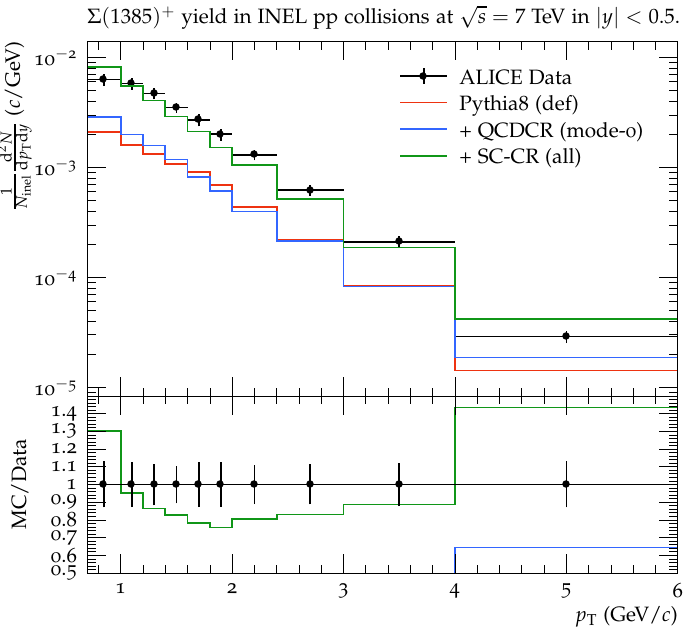}
\end{subfigure}%
\begin{subfigure}{.43\textwidth}
  \includegraphics[width=\linewidth]{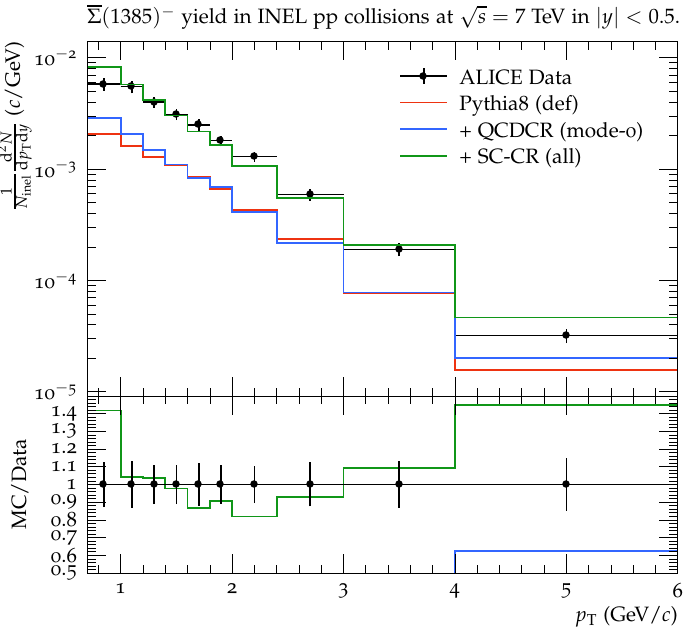}
\end{subfigure}

\begin{subfigure}{.45\textwidth}
  \includegraphics[width=\linewidth]{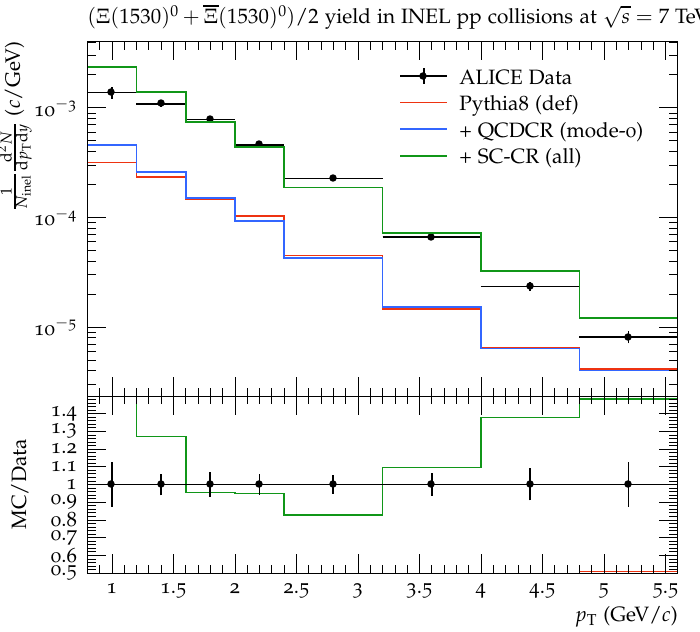}
\end{subfigure}%
\caption{\textit{Top}: $p_\perp$ differential yield of $\Sigma^+$
  (left) and $\Bar{\Sigma}^-$ (right). \textit{Bottom}:
  $(\Xi^0 + \Bar{\Xi}^0)/2$ yield as a function of $p_\perp$. The
  results are from the ALICE experiment \cite{ALICE:2014zxz} for $\pp$
  collisions at 7~TeV and for mid-rapidity ($\mid$y$\mid$ $<
  0.5$). The experimental results are compared with \pythia default,
  \qcdcr (mode-0), and spatially constrained \qcdcr with all the new
  changes from this work, which are shown as red, blue, and green
  lines respectively.}
\label{fig:res_pp7}
\end{figure*}

Since we have forcibly increased the overall strangeness rate in the \pythia
string fragmentation, it is important to check that what we have done
is not unreasonable. In figure \ref{fig:res_pp7} we therefore show
$p_\perp$ distribution for $\Sigma^+$, $\Bar{\Sigma}^-$, and
$(\Xi^0 + \Bar{\Xi}^0)/2$ baryons respectively. The ALICE experiment
\cite{ALICE:2014zxz} results for $\pp$ collisions at $\sqrt{s}=7$~TeV
are used here\footnote{The plots were generated using the
  \setting{ALICE\_2014\_I1300380} routine in Rivet.}. The measurements
are reported for inelastic collisions and for the particles in the
mid-rapidity region ($\mid$y$\mid$ $< 0.5$).

Comparing the default \pythia with and without \qcdcr, it is clear
that the junction reconnections do not contribute much to strange
baryon. Instead, the main production mechanism is diquark breakups in
the string fragmentation. We can therefore conclude that the main
effect when looking at the changes we have done here is the
enhancement of strange (di-)quarks in the string breakups. It can be
argued that our enhancement is a bit high, but it is clearly not
completely unreasonable.

\subsection{Charmed baryon production in $\pp$ collisions}

We now turn to the charmed baryons and will start with
$\Lambda_c$, where we know that the \qcdcr model does a reasonable
job. Looking back at figure \ref{fig:Xic_tests}, we see that our
changes increase the $\Xi^0_c$ rate substantially, and one can fear
that this is compensated by a decrease of $\Lambda_c$.

\begin{figure*}
\centering
\begin{subfigure}{.5\textwidth}
  \includegraphics[width=\linewidth]{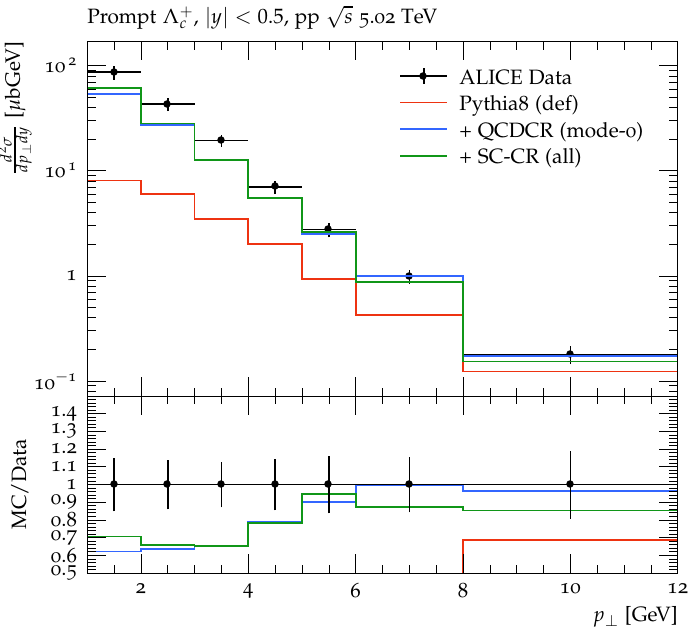}
\end{subfigure}%
\begin{subfigure}{.5\textwidth}
  \includegraphics[width=\linewidth]{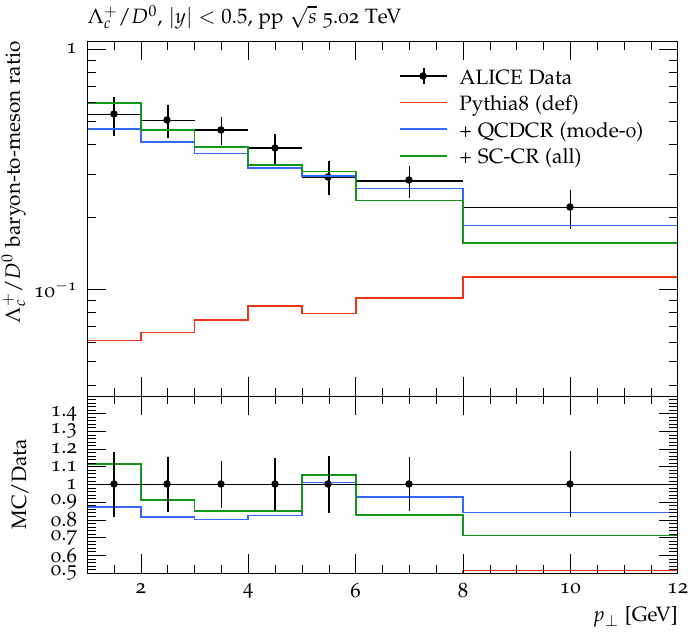}
\end{subfigure}

\caption{Prompt $\Lambda^+_c$ distribution as function of $p_\perp$ on
  the left. The baryon-to-meson ratio for $\Lambda_c^+ /D^0$ as a
  function of $p_\perp$ on the right. The $\pp$ collision at 5.02~TeV
  results are from the ALICE experiment \cite{ALICE:2020wfu}. The
  coloured lines represent the same setups as in figure
  \ref{fig:res_pp7}.}
\label{fig:Lc_res}
\end{figure*}

To check this, we show in figure \ref{fig:Lc_res} the $p_\perp$
distribution of the prompt $\Lambda^+_c$ baryons and
$\Lambda_c^+ /D^0$ ratio is compared with the ALICE data
\cite{ALICE:2020wfu} (using the same rivet
routine as in figure \ref{fig:Xic_tests}). Clearly, we maintain a
good description of the $\Lambda^+_c$ cross section and
$\Lambda_c^+ /D^0$ ratio, even after all the new changes we have
introduced in this work.

We note that the \qcdcr, both with and without our changes, gives more
enhancement for $\Lambda_c^+$ for low $p_\perp$, as seen both for
the yield and for the ratio to the $D^0$ yield. The reason for this
is that most strings in an event are fairly parallel to the beam,
connecting low-$p_\perp$ partons produced by MPI. So the largest
chance to get baryons from junction reconnections is from two or three
dipoles from such strings along the beam direction, which then results
in low-$p_\perp$ baryons.

The effect is less visible for the strange baryons in
figure~\ref{fig:res_pp7} since the relative contribution from junction
reconnection is smaller but it is still reflected in a small increase of
small $p_\perp$ for \qcdcr.

\begin{figure*}
\centering
\begin{subfigure}{.43\textwidth}
  \includegraphics[width=\linewidth]{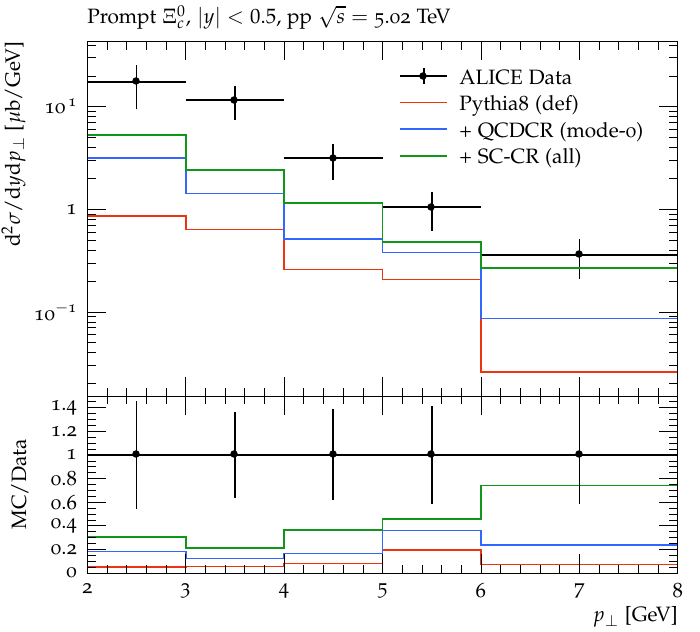}
\end{subfigure}%
\begin{subfigure}{.43\textwidth}
  \includegraphics[width=\linewidth]{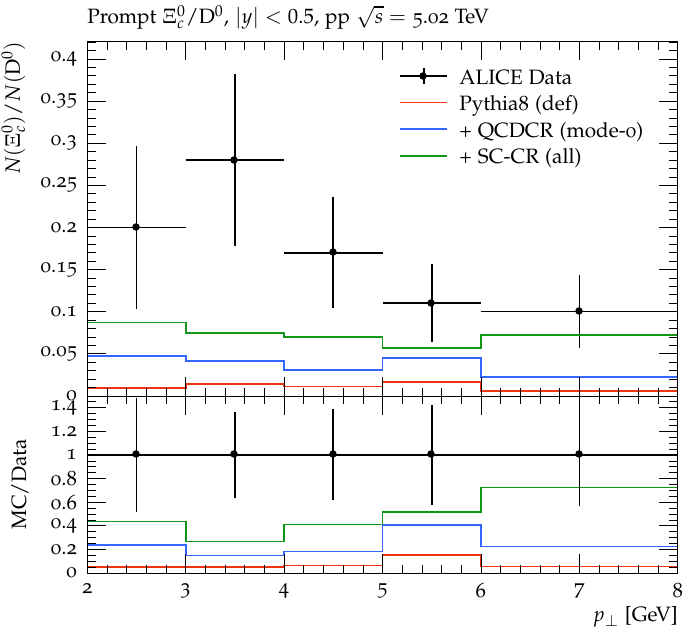}
\end{subfigure}

\begin{subfigure}{.45\textwidth}
  \includegraphics[width=\linewidth]{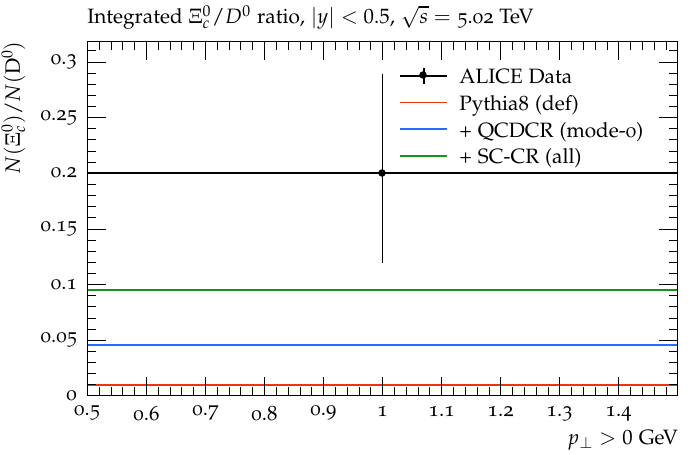}
\end{subfigure}%
\caption{ \textit{Top Left}: The prompt $\Xi^0_c$ cross section as a
  function of $p_\perp$. \textit{Top Right}: The $\Xi^0_c /D^0$ ratio
  as a function of $p_\perp$. \textit{Bottom}: The integrated
  $\Xi^0_c /D^0$ ratio for all $p_\perp >0$. The data is from ALICE
  experiment \cite{ALICE:2021psx} $\pp$ collisions at
  $\sqrt{s} = 5.02$~TeV. The coloured lines represent the same setups
  as in figure \ref{fig:res_pp7}.}
\label{fig:Xic_res}
\end{figure*}

In figure \ref{fig:Xic_res}, we then show the corresponding comparison
for $p_\perp$ distribution of the prompt $\Xi^0_c$ baryons and the
$\Xi^0_c / D^0$ ratio results obtained at the ALICE experiment
\cite{ALICE:2021psx} for $\pp$ collisions at $\sqrt{s}$=5.02~TeV. The
cross section distribution basically shows the same thing that we
previously showed in figure \ref{fig:Xic_tests}, where the overall
yield for \pythia is far below the data while adding \qcdcr brings it
closer, and with our changes even more so.

The $\Xi^0_c / D^0$ ratio is arguably more relevant for assessing
our changes, since the overall (perturbatively modelled) charm rate is
factored out, and only the change in the non-perturbative modelling is
important. Both for the $p_\perp$ distribution and the integrated
ratio our changes actually come quite close to the data (note that
there is a linear scale for the ratios here). We note that for the
$p_\perp$ shape, the data has a tendency to decrease a bit for the
lowest $p_\perp$ bin, while the \qcdcr model, with and without our
changes, seems to continue to rise, mirroring the behaviour in the figure
\ref{fig:Lc_res}.

\begin{figure*}
\centering
\begin{subfigure}{.5\textwidth}
  \includegraphics[width=\linewidth]{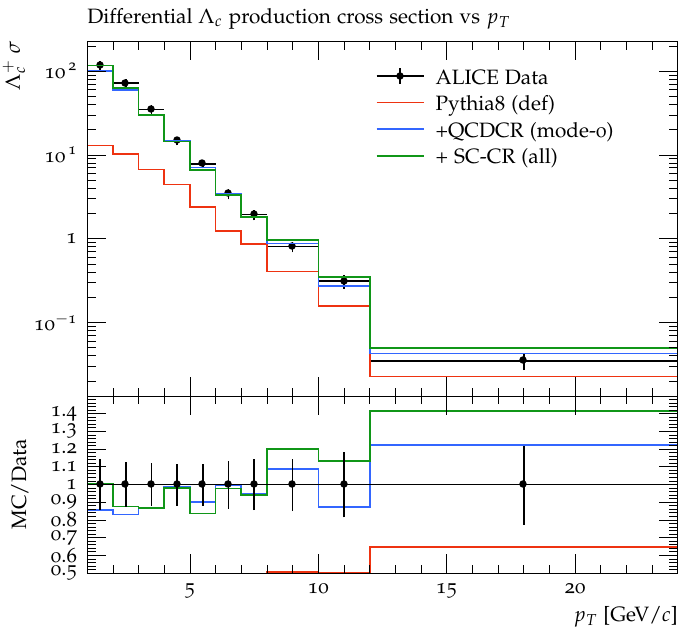}
\end{subfigure}%
\begin{subfigure}{.5\textwidth}
  \includegraphics[width=\linewidth]{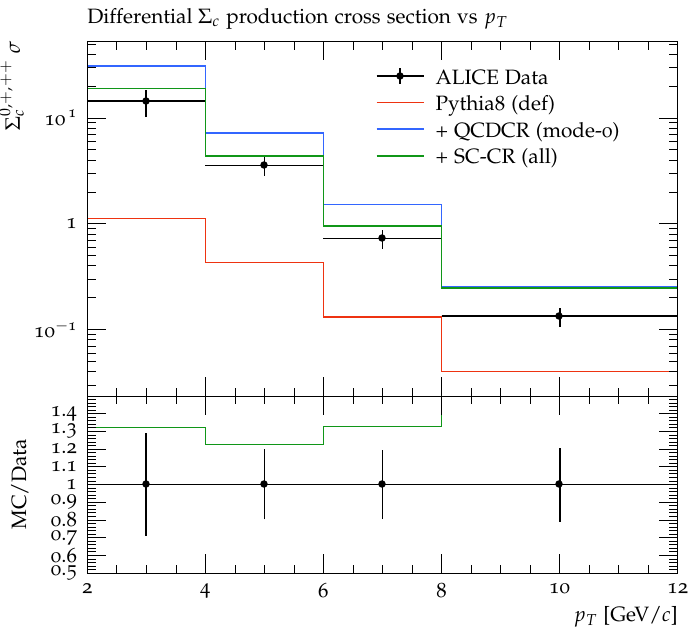}
\end{subfigure}
\caption{$p_\perp$ differential production cross section of
  $\Lambda_c^+$ on the left and of $\Sigma_c^{0,+,++}$ on the right.
  The data is from the ALICE experiment for pp collisions at 13~TeV
  \cite{ALICE:2021rzj}. The coloured lines represent the same setups
  as in figure \ref{fig:res_pp7}.}
\label{fig:res1_pp13}
\end{figure*}

\begin{figure*}
\centering
\begin{subfigure}{.5\textwidth}
  \includegraphics[width=\linewidth]{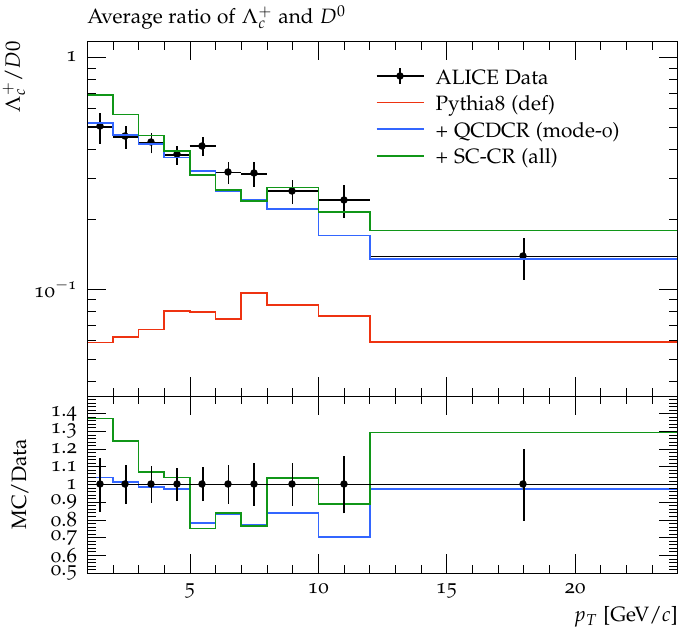}
\end{subfigure}%
\begin{subfigure}{.5\textwidth}
  \includegraphics[width=\linewidth]{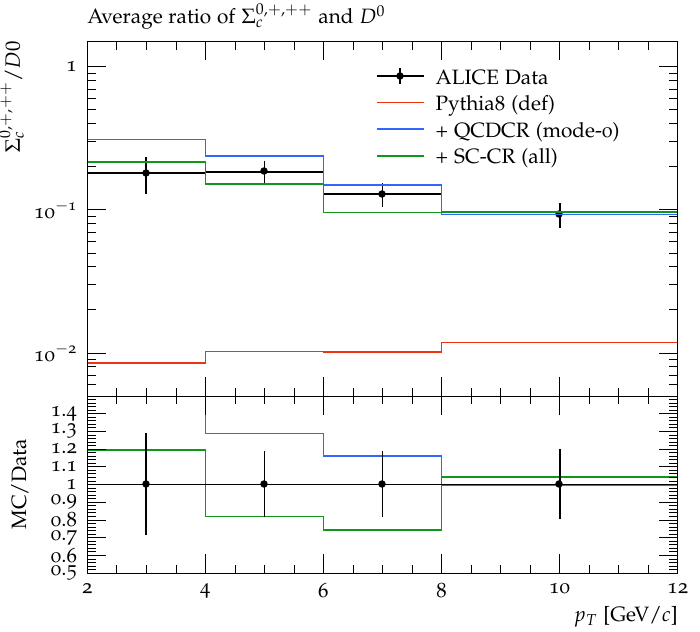}
\end{subfigure}
\caption{Baryon-to-meson ratio for $\Lambda_c^+ /D^0$ on the left and
  $\Sigma_c^{0,+,++} /D^0$ on the right.  The data is from the ALICE
  experiment for pp collisions at 13~TeV \cite{ALICE:2021rzj}. The
  red, blue, and green lines are the same as in the figure
  \ref{fig:res1_pp13}. The coloured lines represent the same setups
  as in figure \ref{fig:res_pp7}.}
\label{fig:res2_pp13}
\end{figure*}

\begin{figure*}
\centering
\begin{subfigure}{.5\textwidth}
  \includegraphics[width=\linewidth]{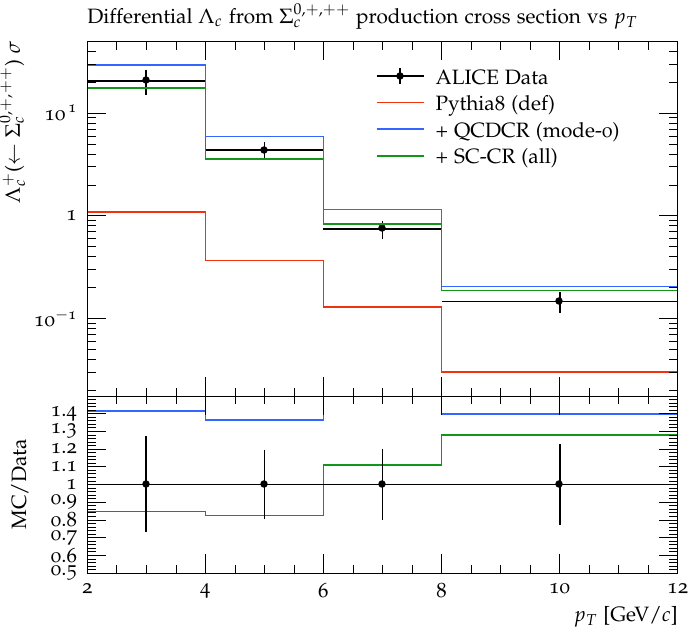}
\end{subfigure}%
\begin{subfigure}{.5\textwidth}
  \includegraphics[width=\linewidth]{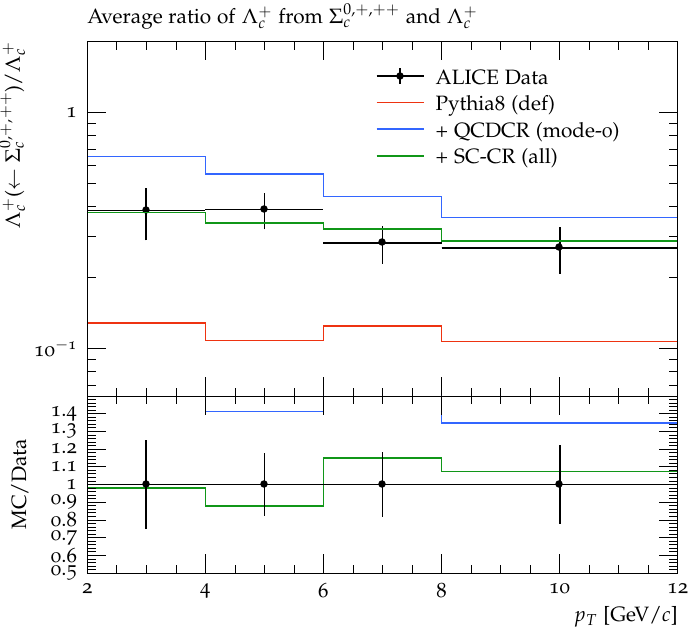}
\end{subfigure}
\caption{\textit{Left}: $p_\perp$ differential production
  cross-section of $\Lambda_c^+$ from $\Sigma_c^{0,+,++}$
  decays. \textit{Right}: Ratio of $\Lambda_c^+$ from
  $\Sigma_c^{0,+,++}$ decays to the total $\Lambda_c^+$ as a function of
  $p_\perp$. The data is from the ALICE experiment for pp collisions
  at 13~TeV \cite{ALICE:2021rzj}.}
\label{fig:res3_pp13}
\end{figure*}

Last year, the ALICE collaboration presented results
\cite{ALICE:2021rzj} \footnote{The analysis is implemented in the
  Rivet routine \setting{ALICE\_2022\_I1868463}.} also for $\Sigma_c$ baryons in \pp collision, this time using data from the LHC run 2 at $\sqrt{s}$= 13~TeV.  Figure \ref{fig:res1_pp13} shows a differential production cross-section for $\Lambda_c^+$ on the left and $\Sigma_c^{0,+,++}$ on the right as a function of $p_\perp$, and in
figure \ref{fig:res2_pp13} the same is shown as a ratio to the $D^0$
cross section.  We can clearly see that the modification of the \qcdcr
model done in this paper not only maintains the $\Lambda_c^+$
description but also controls the $\Sigma_c^{0,+,++}$ production rate
in \pythia.  Finally in figure \ref{fig:res3_pp13}, we show that due
to the reduced $\Sigma_c^{0,+,++}$ production cross-section, the
fraction of $\Lambda_c^+$ coming from $\Sigma_c^{0,+,++}$ decays, and
the ratio to the inclusive $\Lambda_c^+$ both are improved by our
modifications to the \qcdcr model.

From our changes to the \qcdcr, the one mainly influencing the $\Sigma_c$ rate is the change in the parameter controlling the diquark formation in the joining of the smallest junction legs in the fragmentation (see section \ref{sec:juncfrag}). Increasing the probability for a charmed diquark to be in a spin-1 rather than a spin-0 state, means that $\Sigma_c^\star$ states are favoured over the $\Sigma_c$ ones in the subsequent fragmentation of the largest leg. As mentioned in section \ref{sec:juncfrag} these parameters were previously completely unconstrained by data and in \cite{Bierlich:2022pfr}, the authors described the chosen default values as guesswork. In \qcdcr (mode-0) the values were set to the same, rather low, value for all quark types, but in our change, we decided to keep the default ones which are higher and dependent on the heavy quark mass. That the probability should be mass dependent is reasonable since the mass splitting between the spin-1 and spin-0 state should be smaller when heavier quarks are involved. (See, \eg, \cite{Bierlich:2022vdf} for a discussion on this). Thanks to ALICE we now have data \cite{ALICE:2021rzj} that can actually constrain this parameter. Here also we notice that the $\Lambda_c^+ /D^0$ ratio for low
$p_\perp$ is increased.

\subsection{\boldmath $\pPb$ collisions}

\begin{figure*}
\centering
\begin{subfigure}{.5\textwidth}
  \includegraphics[width=\linewidth]{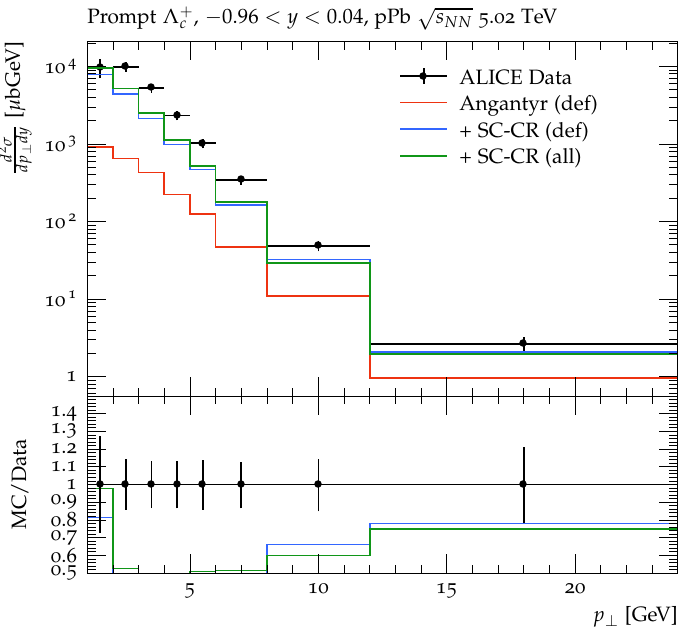}
\end{subfigure}%
\begin{subfigure}{.5\textwidth}
  \includegraphics[width=\linewidth]{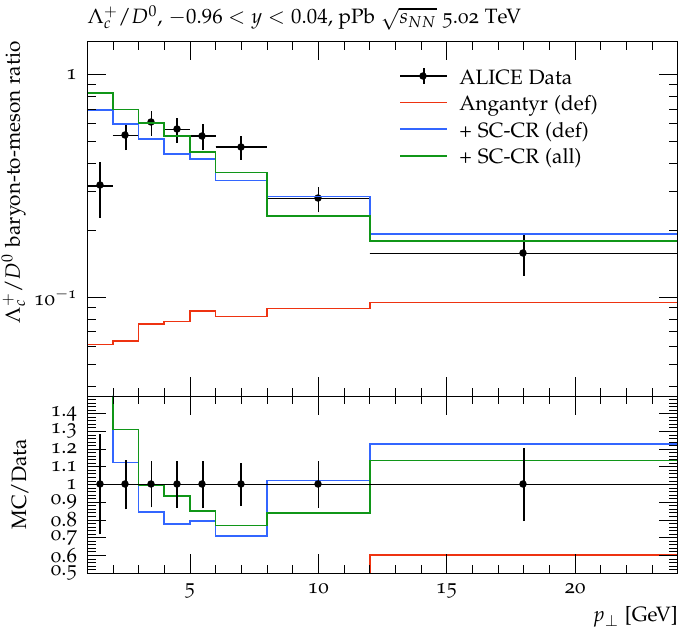}
\end{subfigure}

\caption{Prompt $\Lambda^+_c$ distribution as a function of $p_\perp$
  on the left. The baryon-to-meson ratio, $\Lambda_c^+ /D^0$, as a
  function of $p_\perp$ on the right. The red, blue, and green lines
  are \angantyr default, \angantyr new tune with global CR as SC-CR
  (def), and the changes we have made in this work in the global CR as
  SC-CR (all) respectively. The $pPb$ collision at 5.02~TeV results
  are from the ALICE experiment \cite{ALICE:2020wfu}. (Note that the
  rapidity region, $-0.96<y<0.04$ is given in the collision rest
  frame, and corresponds to the central, $\mid\eta\mid<0.5$ region in the
  laboratory frame.)}
\label{fig:Lc_resPbp}
\end{figure*}

With a reasonable description of charmed baryon production in \pp
collisions, we can now use the \angantyr model to extrapolate our
results in heavy-ion collisions. In a previous publication, we have
shown that the colour reconnection between dipoles in \qcdcr can be constrained by introducing a cut in the transverse separation between dipoles.
By adjusting the value of this cut we can allow for a global colour reconnection between sub-collisions in heavy-ion
collisions and still retain a reasonable description of hadron
multiplicities. Since the charmed baryon production has been shown to
be a sensitive probe into how the junction reconnections in the model
behave, we can now see in more detail if our extrapolation to heavy-ion collisions is reasonable.

In figure \ref{fig:Lc_resPbp} we show \pPb results from the ALICE experiment
at $\sqrt{s_{NN}}$= 5.02~TeV for the $\Lambda_c^+$ cross section, and
the ratio of this w.r.t.\ $D^0$ cross section, in comparison with
\angantyr model. As expected, the default \angantyr, with colour
reconnections only within each sub-collision separately, severely
underestimated the rate of $\Lambda_c^+$. Adding our spatially constrained version of the \qcdcr model improves the description of data significantly, although the $\Lambda_c$ cross section is still somewhat underestimated. Adding the changes introduced in this paper, however, does not influence the result much. This was to be expected, since also in \pp the effect on $\Lambda_c^+$ was minor.

We can see that for low $p_\perp$ the model fails to reproduce
the behaviour of the data. This is best seen in the ratio to $D^0$, where
our model completely shows no sign of reduction of the ratio at small
$p_\perp$. Also this could be expected, as we had also seen
indications of this in \pp collisions above.

\section{Conclusion}

This paper has shown, for the first time, the effect of applying a modern colour reconnection model to a heavy ion collision, in order to better describe baryon yields. We have shown that the production rates of $\Lambda_c^+$ are dramatically improved in \pPb collisions using the \qcdcr model, which has previously worked well in \pp collisions. We also show that the diquark formation in the joining of the junction legs influences the spin-dependent baryon production, and we require experimental data similar to ALICE \cite{ALICE:2021rzj} to constrain the parameter in \pythia.

Heavy quarks can only be produced in hard scattering or in a parton shower mechanism in \pythia.
We show that the application of colour algebra in the \qcdcr model allows junction formation by connecting three colour dipoles in a junction point. These junctions contribute significantly to baryon production.

We show that for a heavy quark connected to a junction the $\lambda$-measure used in the \qcdcr model should be improved.
Usually, the $\lambda$-measure calculates the logarithm of the energy of the dipole in the junction rest frame.
But if the dipole contains a heavy quark then often the invariant mass of the quark has a non-negligible contribution to the energy of the dipole.
Therefore the rapidity span of the heavy quark from the junction point in the junction rest frame should be used as the $\lambda$-measure for such a dipole.

Moreover, when a heavy quark dipole is directly connected to the junction point, the system often fails to obtain a junction rest frame.
If the momentum of the heavy quark is low, it is possible that the string piece between the junction and the heavy quark collapses to zero.
Thus the heavy quark is directly connected by two strings to the lighter quarks.
We show that under such a scenario fragmenting the junction system in the rest frame of the heaviest quark is a good choice.

During the fragmentation of the junction system, the convention is to calculate the energy in every leg and start fragmenting the junction system from the lowest energy leg.
Here again, we show that the choice of the new $\lambda$-measure should be the scalar value of the momentum instead of the energy because we should avoid counting the invariant mass of the quarks as the potential energy available in the junction leg.

We notice that apart from the two modifications in the junction formation during CR and junction fragmentation during hadronization, we need strange quarks as many of the heavy baryons contain strange quarks.
We have a rope hadronization model, which contributes to the strangeness enhancement in \pythia.
The junction topologies are complex and the string-string interactions in rope hadronization haven't been implemented for junction configurations.
Hence we have compensated it by increasing the string fragmentation probability for the strange quarks.

The charm and bottom quark masses are the other parameters we changed in this paper.
To enhance the charm and bottom quark production in the first place we decided to use slightly lower mass values within the proposed mass ranges for the respective quarks.

All these changes together helped us to improve the \pythia description for $\Xi$, $\Sigma$, $\Xi_c$, and $\Sigma_c$ baryon production rates.
We also managed to keep a good description of the $\Lambda^+_c$ and $\Lambda^+_c /D^0$ for different collision energies in $\pp$ collisions.
For the first time, we show the $\Lambda^+_c$ and $\Lambda^+_c /D^0$ results in $pPb$ collisions.
The results are generated with the global CR in \angantyr and with the changes we introduced in this paper, and they show a visible improvement over the default \angantyr setup.

At this stage, it is also important to note that the increased strange quark fragmentation should be replaced with the appropriate treatment of the rope hadronization model.
We may be required to retune some of the parameters because the production of light baryon and other hyperons has not been tracked against the new changes we have made in this work.
Moreover, so far we have applied the $\lambda$-measure correction only to the junction formation, but a similar correction should also be applied to the ``swing'' CR between two dipoles.
We hope that the complete treatment of $\lambda$-measure correction will affect the Quarkonium production.
Hence exploring the possibility of reproducing Quarkonia suppression in heavy-ion collisions is one of the tasks for future work.

\bmhead{Acknowledgments}

\label{sec:acknowledgements}

We would like to thank Torbj\"orn Sj\"ostrand for useful discussions.

This work was funded in part by the Knut and Alice Wallenberg
Foundation, contract number 2017.0036, Swedish Research Council,
contracts numbers 2016-03291 and 2020-04869, in part by the European
Research Council (ERC) under the European Union’s Horizon 2020
research and innovation programme, grant agreement No. 668679, and in
part by the MCnetITN3 H2020 Marie Curie Initial Training Network,
contract 722104.



\bibliography{main}

\begin{thebibliography}{10}
\providecommand{\url}[1]{{#1}}
\providecommand{\urlprefix}{URL }
\providecommand{\doi}[1]{\url{https://doi.org/#1}}
\bibcommenthead

\bibitem{Collins:1985gm}
J.C. Collins, D.E. Soper, G.F. Sterman, {Heavy Particle Production in High-Energy Hadron Collisions}.
\newblock Nucl. Phys. B \textbf{263}, 37 (1986).
\newblock \doi{10.1016/0550-3213(86)90026-X}

\bibitem{Collins:1989gx}
J.C. Collins, D.E. Soper, G.F. Sterman, {Factorization of Hard Processes in QCD}.
\newblock Adv. Ser. Direct. High Energy Phys. \textbf{5}, 1--91 (1989).
\newblock \doi{10.1142/9789814503266_0001}.
\newblock {\href{https://arxiv.org/abs/hep-ph/0409313}{{arXiv:hep-ph/0409313}}}

\bibitem{Nason:1987xz}
P.~Nason, S.~Dawson, R.K. Ellis, {The Total Cross-Section for the Production of Heavy Quarks in Hadronic Collisions}.
\newblock Nucl. Phys. B \textbf{303}, 607--633 (1988).
\newblock \doi{10.1016/0550-3213(88)90422-1}

\bibitem{Nason:1989zy}
P.~Nason, S.~Dawson, R.K. Ellis, {The One Particle Inclusive Differential Cross-Section for Heavy Quark Production in Hadronic Collisions}.
\newblock Nucl. Phys. B \textbf{327}, 49--92 (1989).
\newblock \doi{10.1016/0550-3213(89)90286-1}.
\newblock [Erratum: Nucl.Phys.B 335, 260--260 (1990)]

\bibitem{Mangano:1991jk}
M.L. Mangano, P.~Nason, G.~Ridolfi, {Heavy quark correlations in hadron collisions at next-to-leading order}.
\newblock Nucl. Phys. B \textbf{373}, 295--345 (1992).
\newblock \doi{10.1016/0550-3213(92)90435-E}

\bibitem{Helenius:2018uul}
I.~Helenius, H.~Paukkunen, {Revisiting the D-meson hadroproduction in general-mass variable flavour number scheme}.
\newblock JHEP \textbf{05}, 196 (2018).
\newblock \doi{10.1007/JHEP05(2018)196}.
\newblock {\href{https://arxiv.org/abs/1804.03557}{{arXiv:1804.03557}}} {[hep-ph]}

\bibitem{Cacciari:1998it}
M.~Cacciari, M.~Greco, P.~Nason, {The P(T) spectrum in heavy flavor hadroproduction}.
\newblock JHEP \textbf{05}, 007 (1998).
\newblock \doi{10.1088/1126-6708/1998/05/007}.
\newblock {\href{https://arxiv.org/abs/hep-ph/9803400}{{arXiv:hep-ph/9803400}}}

\bibitem{Cacciari:2012ny}
M.~Cacciari, S.~Frixione, N.~Houdeau, M.L. Mangano, P.~Nason, G.~Ridolfi, {Theoretical predictions for charm and bottom production at the LHC}.
\newblock JHEP \textbf{10}, 137 (2012).
\newblock \doi{10.1007/JHEP10(2012)137}.
\newblock {\href{https://arxiv.org/abs/1205.6344}{{arXiv:1205.6344}}} {[hep-ph]}

\bibitem{Belle:2005mtx}
R.~Seuster, et~al., {Charm hadrons from fragmentation and B decays in e+ e- annihilation at s**(1/2) = 10.6-GeV}.
\newblock Phys. Rev. D \textbf{73}, 032,002 (2006).
\newblock \doi{10.1103/PhysRevD.73.032002}.
\newblock {\href{https://arxiv.org/abs/hep-ex/0506068}{{arXiv:hep-ex/0506068}}}

\bibitem{Kneesch:2007ey}
T.~Kneesch, B.A. Kniehl, G.~Kramer, I.~Schienbein, {Charmed-meson fragmentation functions with finite-mass corrections}.
\newblock Nucl. Phys. B \textbf{799}, 34--59 (2008).
\newblock \doi{10.1016/j.nuclphysb.2008.02.015}.
\newblock {\href{https://arxiv.org/abs/0712.0481}{{arXiv:0712.0481}}} {[hep-ph]}

\bibitem{WA82:1993ghz}
M.~Adamovich, et~al., {Study of D+ and D- Feynman's x distributions in pi- nucleus interactions at the SPS}.
\newblock Phys. Lett. B \textbf{305}, 402--406 (1993).
\newblock \doi{10.1016/0370-2693(93)91074-W}

\bibitem{ALICE:2020wfu}
S.~Acharya, et~al., {$\Lambda^+_c$ Production and Baryon-to-Meson Ratios in pp and p-Pb Collisions at $\sqrt {s_{NN}}$=5.02\,\,TeV at the LHC}.
\newblock Phys. Rev. Lett. \textbf{127}(20), 202,301 (2021).
\newblock \doi{10.1103/PhysRevLett.127.202301}.
\newblock {\href{https://arxiv.org/abs/2011.06078}{{arXiv:2011.06078}}} {[nucl-ex]}

\bibitem{ALICE:2021dhb}
S.~Acharya, et~al., {Charm-quark fragmentation fractions and production cross section at midrapidity in pp collisions at the LHC}.
\newblock Phys. Rev. D \textbf{105}(1), L011,103 (2022).
\newblock \doi{10.1103/PhysRevD.105.L011103}.
\newblock {\href{https://arxiv.org/abs/2105.06335}{{arXiv:2105.06335}}} {[nucl-ex]}

\bibitem{ALICE:2021rzj}
S.~Acharya, et~al., {Measurement of Prompt D$^{0}$, $\Lambda_{c}^{+}$, and $\Sigma_{c}^{0,++}$(2455) Production in Proton\textendash{}Proton Collisions at $\sqrt s$ = 13\,\,TeV}.
\newblock Phys. Rev. Lett. \textbf{128}(1), 012,001 (2022).
\newblock \doi{10.1103/PhysRevLett.128.012001}.
\newblock {\href{https://arxiv.org/abs/2106.08278}{{arXiv:2106.08278}}} {[hep-ex]}

\bibitem{ALICE:2023jgm}
{Exploring the non-universality of charm hadronisation through the measurement of the fraction of jet longitudinal momentum carried by $\Lambda_{\rm c}^+$ baryons in pp collisions}  (2023).
\newblock {\href{https://arxiv.org/abs/2301.13798}{{arXiv:2301.13798}}} {[nucl-ex]}

\bibitem{Bierlich:2022pfr}
C.~Bierlich, et~al., {A comprehensive guide to the physics and usage of PYTHIA 8.3}  (2022).
\newblock \doi{10.21468/SciPostPhysCodeb.8}.
\newblock {\href{https://arxiv.org/abs/2203.11601}{{arXiv:2203.11601}}} {[hep-ph]}

\bibitem{Norrbin:1998bw}
E.~Norrbin, T.~Sjostrand, {Production mechanisms of charm hadrons in the string model}.
\newblock Phys. Lett. B \textbf{442}, 407--416 (1998).
\newblock \doi{10.1016/S0370-2693(98)01244-1}.
\newblock {\href{https://arxiv.org/abs/hep-ph/9809266}{{arXiv:hep-ph/9809266}}}

\bibitem{Norrbin:2000zc}
E.~Norrbin, T.~Sjostrand, {Production and hadronization of heavy quarks}.
\newblock Eur. Phys. J. C \textbf{17}, 137--161 (2000).
\newblock \doi{10.1007/s100520000460}.
\newblock {\href{https://arxiv.org/abs/hep-ph/0005110}{{arXiv:hep-ph/0005110}}}

\bibitem{Andersson:1983ia}
B.~Andersson, G.~Gustafson, G.~Ingelman, T.~Sjostrand, {Parton Fragmentation and String Dynamics}.
\newblock Phys. Rept. \textbf{97}, 31--145 (1983).
\newblock \doi{10.1016/0370-1573(83)90080-7}

\bibitem{QCDCR}
J.R. Christiansen, P.Z. Skands, {String Formation Beyond Leading Colour}.
\newblock JHEP \textbf{08}, 003 (2015).
\newblock \doi{10.1007/JHEP08(2015)003}.
\newblock {\href{https://arxiv.org/abs/1505.01681}{{arXiv:1505.01681}}} {[hep-ph]}

\bibitem{ALICE:2021bli}
S.~Acharya, et~al., {Measurement of the Cross Sections of $\Xi^0_{c}$ and $\Xi^+_{c}$ Baryons and of the Branching-Fraction Ratio BR($\Xi^0_{c} \rightarrow \Xi^-{e}^+\nu_{ e}$)/BR($\Xi^0_{c} \rightarrow \Xi^-\pi^+$) in pp collisions at 13 TeV}.
\newblock Phys. Rev. Lett. \textbf{127}(27), 272,001 (2021).
\newblock \doi{10.1103/PhysRevLett.127.272001}.
\newblock {\href{https://arxiv.org/abs/2105.05187}{{arXiv:2105.05187}}} {[nucl-ex]}

\bibitem{ALICE:2021psx}
S.~Acharya, et~al., {Measurement of the production cross section of prompt $ {\Xi}_{\mathrm{c}}^0 $ baryons at midrapidity in pp collisions at $ \sqrt{s} $ = 5.02 TeV}.
\newblock JHEP \textbf{10}, 159 (2021).
\newblock \doi{10.1007/JHEP10(2021)159}.
\newblock {\href{https://arxiv.org/abs/2105.05616}{{arXiv:2105.05616}}} {[nucl-ex]}

\bibitem{ALICE:2022cop}
{First measurement of $\rm \Omega_c^0$ production in pp collisions at $\sqrt{s}=13$ TeV}  (2022).
\newblock {\href{https://arxiv.org/abs/2205.13993}{{arXiv:2205.13993}}} {[nucl-ex]}

\bibitem{Lonnblad:2023stc}
L.~L\"onnblad, H.~Shah, {A spatially constrained QCD colour reconnection in $\textrm{pp}$, $\textrm{p}A$, and $AA$ collisions in the Pythia8/Angantyr model}.
\newblock Eur. Phys. J. C \textbf{83}(7), 575 (2023).
\newblock \doi{10.1140/epjc/s10052-023-11778-3}.
\newblock {\href{https://arxiv.org/abs/2303.11747}{{arXiv:2303.11747}}} {[hep-ph]}

\bibitem{Angantyr}
C.~Bierlich, G.~Gustafson, L.~L{\"o}nnblad, H.~Shah, {The Angantyr model for Heavy-Ion Collisions in PYTHIA8}.
\newblock JHEP \textbf{10}, 134 (2018).
\newblock \doi{10.1007/JHEP10(2018)134}.
\newblock {\href{https://arxiv.org/abs/1806.10820}{{arXiv:1806.10820}}} {[hep-ph]}

\bibitem{Bierlich:2016smv}
C.~Bierlich, G.~Gustafson, L.~L\"onnblad, {Diffractive and non-diffractive wounded nucleons and final states in pA collisions}.
\newblock JHEP \textbf{10}, 139 (2016).
\newblock \doi{10.1007/JHEP10(2016)139}.
\newblock {\href{https://arxiv.org/abs/1607.04434}{{arXiv:1607.04434}}} {[hep-ph]}

\bibitem{Glauber:1955qq}
R.J. Glauber, {Cross-sections in deuterium at high-energies}.
\newblock Phys. Rev. \textbf{100}, 242--248 (1955).
\newblock \doi{10.1103/PhysRev.100.242}

\bibitem{Miller:2007ri}
M.L. Miller, K.~Reygers, S.J. Sanders, P.~Steinberg, {Glauber modeling in high energy nuclear collisions}.
\newblock Ann. Rev. Nucl. Part. Sci. \textbf{57}, 205--243 (2007).
\newblock \doi{10.1146/annurev.nucl.57.090506.123020}.
\newblock {\href{https://arxiv.org/abs/nucl-ex/0701025}{{arXiv:nucl-ex/0701025}}}

\bibitem{Bozek:2019wyr}
P.~Bo\.zek, W.~Broniowski, M.~Rybczynski, G.~Stefanek, {GLISSANDO 3: GLauber Initial-State Simulation AND mOre..., ver. 3}.
\newblock Comput. Phys. Commun. \textbf{245}, 106,850 (2019).
\newblock \doi{10.1016/j.cpc.2019.07.014}.
\newblock {\href{https://arxiv.org/abs/1901.04484}{{arXiv:1901.04484}}} {[nucl-th]}

\bibitem{Heiselberg:1991is}
H.~Heiselberg, G.~Baym, B.~Blaettel, L.L. Frankfurt, M.~Strikman, {Color transparency, color opacity, and fluctuations in nuclear collisions}.
\newblock Phys. Rev. Lett. \textbf{67}, 2946--2949 (1991).
\newblock \doi{10.1103/PhysRevLett.67.2946}

\bibitem{Blaettel:1993ah}
B.~Blaettel, G.~Baym, L.L. Frankfurt, H.~Heiselberg, M.~Strikman, {Hadronic cross-section fluctuations}.
\newblock Phys. Rev. D \textbf{47}, 2761--2772 (1993).
\newblock \doi{10.1103/PhysRevD.47.2761}

\bibitem{Alvioli:2013vk}
M.~Alvioli, M.~Strikman, {Color fluctuation effects in proton-nucleus collisions}.
\newblock Phys. Lett. B \textbf{722}, 347--354 (2013).
\newblock \doi{10.1016/j.physletb.2013.04.042}.
\newblock {\href{https://arxiv.org/abs/1301.0728}{{arXiv:1301.0728}}} {[hep-ph]}

\bibitem{Sjostrand:1987su}
T.~Sj{\"o}strand, M.~van Zijl, {A Multiple Interaction Model for the Event Structure in Hadron Collisions}.
\newblock Phys. Rev. D \textbf{36}, 2019 (1987).
\newblock \doi{10.1103/PhysRevD.36.2019}

\bibitem{Bierlich:2023ewv}
C.~Bierlich, J.~Wilkinson, J.~Sun, G.~Manca, R.~Granier~de Cassagnac, J.~Otwinowski, {Open charm production cross section from combined LHC experiments in $pp$ collisions at $\sqrt{s} = 5.02$ TeV}  (2023).
\newblock {\href{https://arxiv.org/abs/2311.11426}{{arXiv:2311.11426}}} {[hep-ph]}

\bibitem{LSM}
B.~Andersson, G.~Gustafson, G.~Ingelman, T.~Sj{\"o}strand, Parton fragmentation and string dynamics.
\newblock Phys. Rept. \textbf{97}, 31--145 (1983)

\bibitem{Skands:2014pea}
P.~Skands, S.~Carrazza, J.~Rojo, {Tuning PYTHIA 8.1: the Monash 2013 Tune}.
\newblock Eur. Phys. J. C \textbf{74}(8), 3024 (2014).
\newblock \doi{10.1140/epjc/s10052-014-3024-y}.
\newblock {\href{https://arxiv.org/abs/1404.5630}{{arXiv:1404.5630}}} {[hep-ph]}

\bibitem{Bierlich:2015rha}
C.~Bierlich, J.R. Christiansen, {Effects of color reconnection on hadron flavor observables}.
\newblock Phys. Rev. D \textbf{92}(9), 094,010 (2015).
\newblock \doi{10.1103/PhysRevD.92.094010}.
\newblock {\href{https://arxiv.org/abs/1507.02091}{{arXiv:1507.02091}}} {[hep-ph]}

\bibitem{Bar:popcorn}
B.~Andersson, G.~Gustafson, T.~Sj{\"o}strand, {Baryon Production in Jet Fragmentation and $\Upsilon$ Decay}.
\newblock Phys. Scripta \textbf{32}, 574 (1985).
\newblock \doi{10.1088/0031-8949/32/6/003}

\bibitem{TPCTwoGamma:1985zxy}
H.~Aihara, et~al., {Baryon production in $e^+e^-$ annihilation at $\sqrt{s}=29$ GeV: clusters, diquarks, popcorn?}
\newblock Phys. Rev. Lett. \textbf{55}, 1047 (1985).
\newblock \doi{10.1103/PhysRevLett.55.1047}

\bibitem{Andersson:1988ee}
B.~Andersson, P.~Dahlkvist, G.~Gustafson, {AN INFRARED STABLE MULTIPLICITY MEASURE ON QCD PARTON STATES}.
\newblock Phys. Lett. B \textbf{214}, 604--608 (1988).
\newblock \doi{10.1016/0370-2693(88)90128-1}

\bibitem{ALICE:2016fzo}
J.~Adam, et~al., {Enhanced production of multi-strange hadrons in high-multiplicity proton-proton collisions}.
\newblock Nature Phys. \textbf{13}, 535--539 (2017).
\newblock \doi{10.1038/nphys4111}.
\newblock {\href{https://arxiv.org/abs/1606.07424}{{arXiv:1606.07424}}} {[nucl-ex]}

\bibitem{Bierlich:2014xba}
C.~Bierlich, G.~Gustafson, L.~L\"onnblad, A.~Tarasov, {Effects of Overlapping Strings in pp Collisions}.
\newblock JHEP \textbf{03}, 148 (2015).
\newblock \doi{10.1007/JHEP03(2015)148}.
\newblock {\href{https://arxiv.org/abs/1412.6259}{{arXiv:1412.6259}}} {[hep-ph]}

\bibitem{Bierlich:2022oja}
C.~Bierlich, S.~Chakraborty, G.~Gustafson, L.~L\"onnblad, {Jet modifications from colour rope formation in dense systems of non-parallel strings}.
\newblock SciPost Phys. \textbf{13}(2), 023 (2022).
\newblock \doi{10.21468/SciPostPhys.13.2.023}.
\newblock {\href{https://arxiv.org/abs/2202.12783}{{arXiv:2202.12783}}} {[hep-ph]}

\bibitem{Bierlich:2022ned}
C.~Bierlich, S.~Chakraborty, G.~Gustafson, L.~L\"onnblad, {Strangeness enhancement across collision systems without a plasma}.
\newblock Phys. Lett. B \textbf{835}, 137,571 (2022).
\newblock \doi{10.1016/j.physletb.2022.137571}.
\newblock {\href{https://arxiv.org/abs/2205.11170}{{arXiv:2205.11170}}} {[hep-ph]}

\bibitem{Bierlich:2019rhm}
C.~Bierlich, et~al., {Robust Independent Validation of Experiment and Theory: Rivet version 3}.
\newblock SciPost Phys. \textbf{8}, 026 (2020).
\newblock \doi{10.21468/SciPostPhys.8.2.026}.
\newblock {\href{https://arxiv.org/abs/1912.05451}{{arXiv:1912.05451}}} {[hep-ph]}

\bibitem{ALICE:2014zxz}
B.B. Abelev, et~al., {Production of $\Sigma(1385)^{\pm}$ and $\Xi(1530)^{0}$ in proton-proton collisions at $\sqrt{s}=$ 7 TeV}.
\newblock Eur. Phys. J. C \textbf{75}(1), 1 (2015).
\newblock \doi{10.1140/epjc/s10052-014-3191-x}.
\newblock {\href{https://arxiv.org/abs/1406.3206}{{arXiv:1406.3206}}} {[nucl-ex]}

\bibitem{Bierlich:2022vdf}
C.~Bierlich, S.~Chakraborty, G.~Gustafson, L.~L\"onnblad, {Hyperfine splitting effects in string hadronization}.
\newblock Eur. Phys. J. C \textbf{82}(3), 228 (2022).
\newblock \doi{10.1140/epjc/s10052-022-10172-9}.
\newblock {\href{https://arxiv.org/abs/2201.06316}{{arXiv:2201.06316}}} {[hep-ph]}

\end{thebibliography}


\end{document}